\documentclass[twocolumn,secnumarabic,amssymb, nobibnotes, aps, prl, superscriptaddress]{revtex4-1}

\usepackage{amsmath}
\usepackage{amsfonts}
\usepackage{multirow,wrapfig}
\usepackage{color}
\usepackage{graphicx}
\usepackage{cancel}

\newcommand{\abs}[1]{\ensuremath{\left| #1 \right|}}
\newcommand{\bra}[1]{\ensuremath{\left< #1 \right|}}
\newcommand{\ket}[1]{\ensuremath{\left| #1 \right>}}
\newcommand{\brac}[2]{\ensuremath{\left< #1\left|#2 \right. \right>}}
\newcommand{\avg}[1]{\ensuremath{\left< #1 \right>}}

\newcommand{\h}{\hat}
\newcommand{\be}[0]{\begin{equation}}
\newcommand{\ee}[0]{\end{equation}}
\newcommand{\bea}[0]{\begin{eqnarray}}
\newcommand{\eea}[0]{\end{eqnarray}}

\begin{document}

%\title{Direct Measurement of a 27-Dimensional Orbital Angular Momentum State Vector}%
\title{Direct Measurement of a 27-Dimensional Orbital-Angular-Momentum State Vector}

\author{Mehul Malik}%
\email{mehul.malik@univie.ac.at}
\affiliation{The Institute of Optics, University of Rochester, Rochester, New York 14627, USA}
\affiliation{Institute for Quantum Optics and Quantum Information (IQOQI), Austrian Academy of Sciences, Vienna, Austria}

\author{Mohammad Mirhoseinni}
\affiliation{The Institute of Optics, University of Rochester, Rochester, New York 14627, USA}

\author{Martin P. J. Lavery}
\affiliation{School of Physics and Astronomy, University of Glasgow, Glasgow, United Kingdom}

\author{Jonathan Leach}
\affiliation{School of Engineering \& Physical Sciences, Heriot-Watt University, Edinburgh, United Kingdom}
\affiliation{Department of Physics, University of Ottawa, Ottawa, Ontario, Canada}

\author{Miles J. Padgett}
\affiliation{School of Physics and Astronomy, University of Glasgow, Glasgow, United Kingdom}

\author{Robert W. Boyd}
\affiliation{The Institute of Optics, University of Rochester, Rochester, New York 14627, USA}
\affiliation{Department of Physics, University of Ottawa, Ottawa, Ontario, Canada}

\date{January 18, 2014}%

\begin{abstract}
The measurement of a quantum state poses a unique challenge for experimentalists. Recently, the technique of ``direct measurement" was proposed for characterizing a quantum state in-situ through sequential weak and strong measurements. While this method has been used for measuring polarization states, its real potential lies in the measurement of states with a large dimensionality. Here we show the practical direct measurement of a high-dimensional state vector in the discrete basis of orbital-angular momentum. Through weak measurements of orbital-angular momentum and strong measurements of angular position, we measure the complex probability amplitudes of a pure state with a dimensionality, d=27. Further, we use our method to directly observe the relationship between rotations of a state vector and the relative phase between its orbital-angular-momentum components. Our technique has important applications in high-dimensional classical and quantum information systems, and can be extended to characterize other types of large quantum states.

\end{abstract}

\maketitle

The measurement problem in quantum mechanics has led us to constantly redefine what we mean by a quantum state \cite{Neumann:1955,Wheeler:1983}. The act of measuring a quantum state disturbs it irreversibly, a phenomenon referred to as collapse of the wavefunction. For example, precisely measuring the position of a single photon results in a photon with a broad superposition of momenta. Consequently, no quantum system can be fully characterized through a single measurement. An established method of characterizing a quantum state involves making a diverse set of measurements on a collection of identically prepared quantum states, followed by post-processing of the data. This process, known as quantum state tomography \cite{DAriano:2003tw}, is akin to its classical counterpart of imaging a three-dimensional object using many two-dimensional projections. A novel alternative to tomography was recently demonstrated in which the complex probability amplitude of a pure quantum state was directly obtained as an output of the measurement apparatus, bypassing the complicated post-processing step \cite{Lundeen:2011isa}. In this method, the position of an ensemble of identically prepared photons was weakly measured\cite{Aharonov:1988fk,Dressel:2013ti}, which caused a minimal disturbance to their momentum. A subsequent strong measurement of their momentum revealed all the information necessary to characterize their state in the continuous bases of position and momentum.

%For a simple quantum system such as a polarization qubit, quantum tomography can be similarly visualized as making projections onto different axes of the Poincare sphere in order to localize the state on the sphere \cite{Altepeter:2005dc}. A critical part of any real tomographic process is the analysis that follows this series of measurements --- in order to obtain a physical quantum state, one must use lengthy numerical procedures to search over all the different state possibilities \cite{Banaszek:1999ho}. The time required for this post-processing step scales rather unfavorably with state dimension, and is catastrophically large for multipartite high-dimensional states \cite{Thew:2002tc,Agnew:2011js}.

%211 words

While a careful study comparing direct measurement with quantum state tomography is still lacking, it has already shown promise as a simple and elegant method for characterizing a quantum state in the continuous position-momentum basis \cite{Lundeen:2011isa} as well as the two-dimensional polarization basis \cite{Salvail:2013bo}. Here, we extend this novel technique to characterize a photon in the discrete, unbounded state-space of orbital-angular-momentum (OAM). Photons carrying OAM have been the subject of much recent scientific attention \cite{MolinaTerriza:2007ig,Galvez:2011iq,Yao:2011ve,Fickler:2012hj}. The discrete, high dimensionality of the OAM Hilbert space provides a much larger information capacity for quantum information systems \cite{Groeblacher:2005ec,Malik:2012ka} as compared to the conventionally used two-dimensional state-space of polarization. More significantly, a larger dimensionality allows for an increased tolerance to eavesdropping in quantum key distribution \cite{Bourennane:2002uo}. Photons entangled in OAM \cite{Mair:2001ub,Leach:2010bm} are prime candidates not only for such high capacity, high security communication systems, but also for fundamental tests of quantum mechanics \cite{Dada:2011dn}. Thus, it is essential that fast, accurate, and efficient methods for characterizing such high-dimensional states be developed. 

In this article, we use weak measurements of OAM followed by a strong measurement of angular position to directly measure the complex probability amplitudes of a 27-dimensional state in the OAM basis. In this manner, we are able to obtain both the amplitude and the phase of each OAM component within our state space. In addition, our technique enables us to measure rotations of a state vector in the natural basis of OAM. Rotation of the state by a fixed angle manifests as an OAM mode-dependent phase, illustrating the relationship between the angular momentum operator and rotations in quantum mechanics \cite{Shankar:1994}.

\begin{figure*}[t!]
\centering\includegraphics[scale=.45]{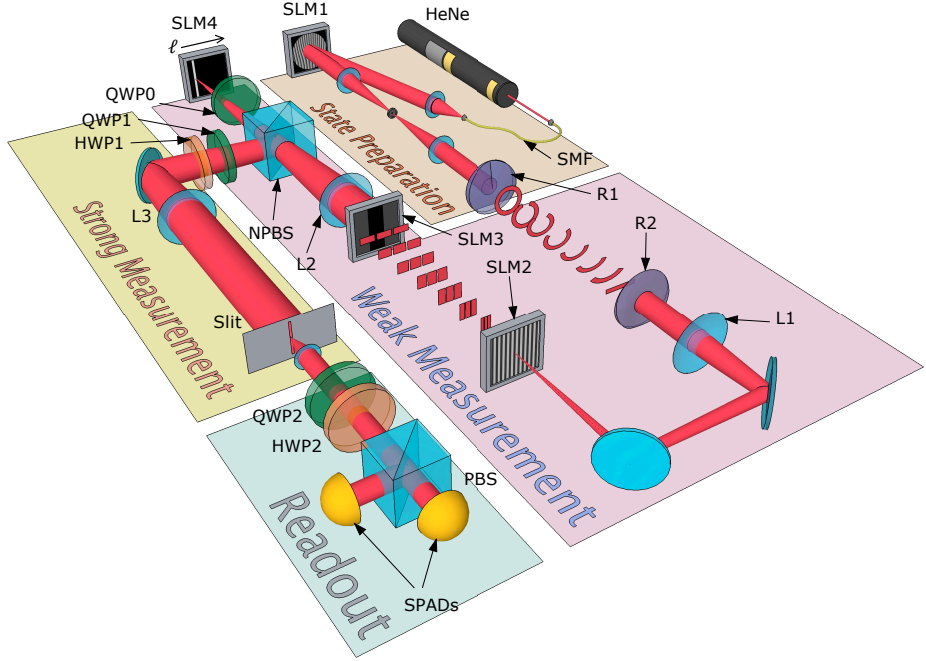}
\caption{Experimental setup for direct measurement of a high-dimensional state vector. State Preparation: A quantum state in an arbitrary superposition of orbital-angular-momentum (OAM) modes is prepared by impressing phase information with a spatial light modulator (SLM1) onto spatially-filtered (SMF) photons from an attenuated HeNe laser. Weak Measurement: A particular OAM mode is weakly projected by rotating its polarization. In order to do so, the OAM modes are first transformed into finite-sized momentum modes by two refractive optical elements made out of PMMA (R1 and R2). Then, a Fourier transform lens (L1) and a fan-out hologram implemented on SLM2 are used to generate three adjacent copies of each momentum mode. The phase between these copies is corrected by SLM3. Another lens (L2) converts these larger momentum modes into well separated position modes at its focus. Finally, a quarter-wave plate (QWP0) used in double-pass with SLM4 are used to rotate the polarization of the OAM mode to be weakly projected. Another quarter-wave and half-wave plate (QWP1 and HWP1) are used to remove any ellipticity introduced by transmission and reflection through the non-polarizing beam splitter (NPBS). Strong measurement: A strong measurement of angular position is performed by Fourier transforming with a lens (L3) and post-selecting state $p=0$ with a 10 $\mu$m slit. Readout: The OAM weak value $\avg{\pi_{\ell}}_{\textrm{w}}$ is obtained by measuring the change in the photon polarization in the linear and circular polarization bases. QWP2, HWP2, a polarizing beam splitter (PBS), and two single-photon avalanche detectors (SPADs) are used for this purpose.}\label{DMsetup}
\end{figure*}  

\section*{Results}

\textbf{Theory of direct measurement in the OAM basis.} We can express the state of our photon as a superposition of states in the OAM basis as

\be \ket{\Psi}=\sum_{\ell}a_{\ell}\ket{\ell},\label{states}\ee

\noindent where $a_{\ell}$ are complex probability amplitudes. In direct analogy to a photon's position and linear momentum, the angular position and OAM of a photon form a discrete Fourier conjugate pair \cite{Yao:2006vc,Jack:2008gx}. Consequently, any OAM basis state $\ket{\ell}$ is mutually unbiased with respect to any angular position basis state $\ket{\theta}$, i.e. their inner product always has the same magnitude. This property allows us to define a strategic quantity $c=\brac{\theta_0}{\ell}/\brac{\theta_0}{\Psi}$, which is constant with respect to $\ket{\ell}$ for $\theta_0=0$. By multiplying our state above by this constant and inserting the identity, we can expand it as 

\be  c \ket{\Psi} = c \sum_{\ell} \ket{\ell}\brac{\ell}{\Psi} = \sum_{\ell} \ket{\ell} \frac{\brac{\theta_0}{\ell}\brac{\ell}{\Psi}}{\brac{\theta_0}{\Psi}}= \sum_{\ell}\avg{\pi_{\ell}}_{\textrm{w}}\ket{\ell}.\ee

\noindent Notice here that we have introduced the quantity $\avg{\pi_{\ell}}_{\textrm{w}}$, which is proportional to the probability amplitude $a_{\ell}$ from Eq. (\ref{states}). This is known as the OAM weak value \cite{Aharonov:1988fk,Dressel:2013ti}, and is equal to the average result obtained by making a weak projection in the OAM basis ($\h{\pi}_{\ell} = \ket{\ell}\bra{\ell}$) followed by a strong measurement in the conjugate basis of angular position ($\theta$). In this manner, the scaled complex probability amplitudes $ca_\ell$ can be directly obtained by measuring the OAM weak value $\avg{\pi_{\ell}}_{\textrm{w}}$ for a finite set of $\ell$. Following this procedure, the constant $c$ can be eliminated by renormalizing the state $\ket{\Psi}$. In order to measure such weak values, we utilize a two-system Hamiltonian where the OAM of a photon is coupled to its polarization, which serves as a measurement pointer \cite{Lundeen:2011isa}. We perform a weak projection of OAM by rotating by a small angle the polarization of the OAM mode to be measured. Following this, a strong measurement of angular position is performed via a post-selection of states with $\theta=\theta_0$. The OAM weak value is read out by measuring the average change in the photon's linear and circular polarization (see Methods section for details).\\

%835 words

\noindent\textbf{Experimental procedure for measuring the OAM weak value.} Performing a weak measurement of OAM at the single photon level is an experimental challenge. In order to do so, we first use an optical geometric transformation in combination with a beam-copying technique to efficiently separate the OAM modes of our photons \cite{Mirhosseini:2013em,Berkhout:2010cb,Lavery:2012hi}. This process is depicted in Fig. \ref{DMsetup} for a single OAM mode. R1 and R2 are custom refractive elements that transform an OAM mode with azimuthal phase variation $e^{i\ell\theta}$ to a momentum mode with position phase variation $e^{i\ell x/a}$. Following a Fourier transform lens (L1), a fan-out hologram implemented on a phase-only spatial light modulator (SLM2) creates three adjacent copies of this momentum mode \cite{Prongue:1992uj,OSullivan:2012gj}. Following another Fourier transform lens (not shown in Fig. \ref{DMsetup}), SLM3 is used to remove a relative phase difference introduced in the beam-copying process between the three copies. The resultant momentum mode is three times the size of the original, while also having three times the phase gradient of the original. A second lens (L2) Fourier transforms this larger momentum mode into a position mode at SLM4. This results in well separated OAM modes ($\ell$) having less than 10\% overlap on average with neighboring modes ($\ell\pm1$) (see Methods section for details). 

The weak projection of an OAM mode is performed by rotating its polarization by an angle $\alpha=\pi/9$ (a strong projection would correspond to $\alpha=\pi/2$). We use SLM4 and a quarter-wave plate (QWP0) in double pass to carry out this polarization rotation \cite{Davis:2000ur}. QWP1 and HWP1 are used to remove any ellipticity introduced by transmission and reflection through the non-polarizing beamsplitter (NPBS). A strong measurement of angular position is performed by a 10 $\mu$m slit placed in the Fourier plane of lens L3. Since the plane of the slit is conjugate to the plane where the OAM modes are spatially separated (SLM4), a measurement of linear position by the slit is equivalent to a measurement of angular position.

%1098 words

%In the next step, we rotate the polarization of the OAM mode to be weakly measured by an angle, $\alpha=\pi/9$ (a strong measurement would correspond to $\alpha=\pi/2$). In contrast to the dynamic method used by Lundeen et al \cite{Lundeen:2011isa} in which they physically moved a half-wave plate (HWP) sliver through the beam, we use a static, programmable technique. A phase-only SLM acts as a variable phase retarder with individually addressable pixels. By sandwiching such an SLM between two quarter-wave plates (QWPs) whose extraordinary axes are aligned at $\pi/4$ radians to the SLM axis, one can rotate the polarization of any part of the beam through an arbitrary angle\cite{Davis:2000ur}. As shown in Fig. \ref{DMsetup}, we use SLM4 and QWP0 (in double pass) to rotate the polarization of the OAM mode to be weakly measured. QWP1 and HWP1 are used to remove any ellipticity introduced by reflection through the non-polarizing beamsplitter (NPBS). A 10 $\mu$m slit is placed in the Fourier plane of lens L3, which constitutes the strong measurement of angular position ($\theta=0$). 

The average change in the photon's linear and circular polarization is proportional to Re$\avg{\pi_{\ell}}_{\textrm{w}}$ and Im$\avg{\pi_{\ell}}_{\textrm{w}}$ respectively \cite{Lundeen:2011isa}. If the initial polarization of the photon is vertical, the OAM weak value is given by

\be\avg{\pi_{\ell}}_{\textrm{w}}=\frac{1}{\sin\alpha}\bigg(\bra{s_\textrm{f}}\h\sigma_1\ket{s_\textrm{f}}-i\bra{s_\textrm{f}}\h\sigma_2\ket{s_\textrm{f}}\bigg),\label{Pauli}\ee

\noindent where $\alpha$ is the rotation angle, $\h{\sigma}_1$ and $\h{\sigma}_2$ are the first and second Pauli operators, and $\ket{s_\textrm{f}}$ is the final polarization state of the photon (see Methods section for detailed derivation of Eq. \ref{Pauli}). In order to measure the expectation values of $\h{\sigma}_1$ and $\h{\sigma}_2$, we transform to the linear and circular polarization bases with QWP2 and HWP2, and measure the difference between orthogonal polarization components with a polarizing beamsplitter (PBS) and two single-photon avalanche detectors (SPADs). In this manner, we directly obtain the scaled complex probability amplitudes $ca_{\ell}$ by scanning the weak measurement through $\ell$ values of $\pm13$. While the size of the OAM state space is unbounded, we are limited to a dimensionality of $d=27$ by our mode transformation technique.\\

\begin{figure}[t!]
\centering\includegraphics[scale=.53]{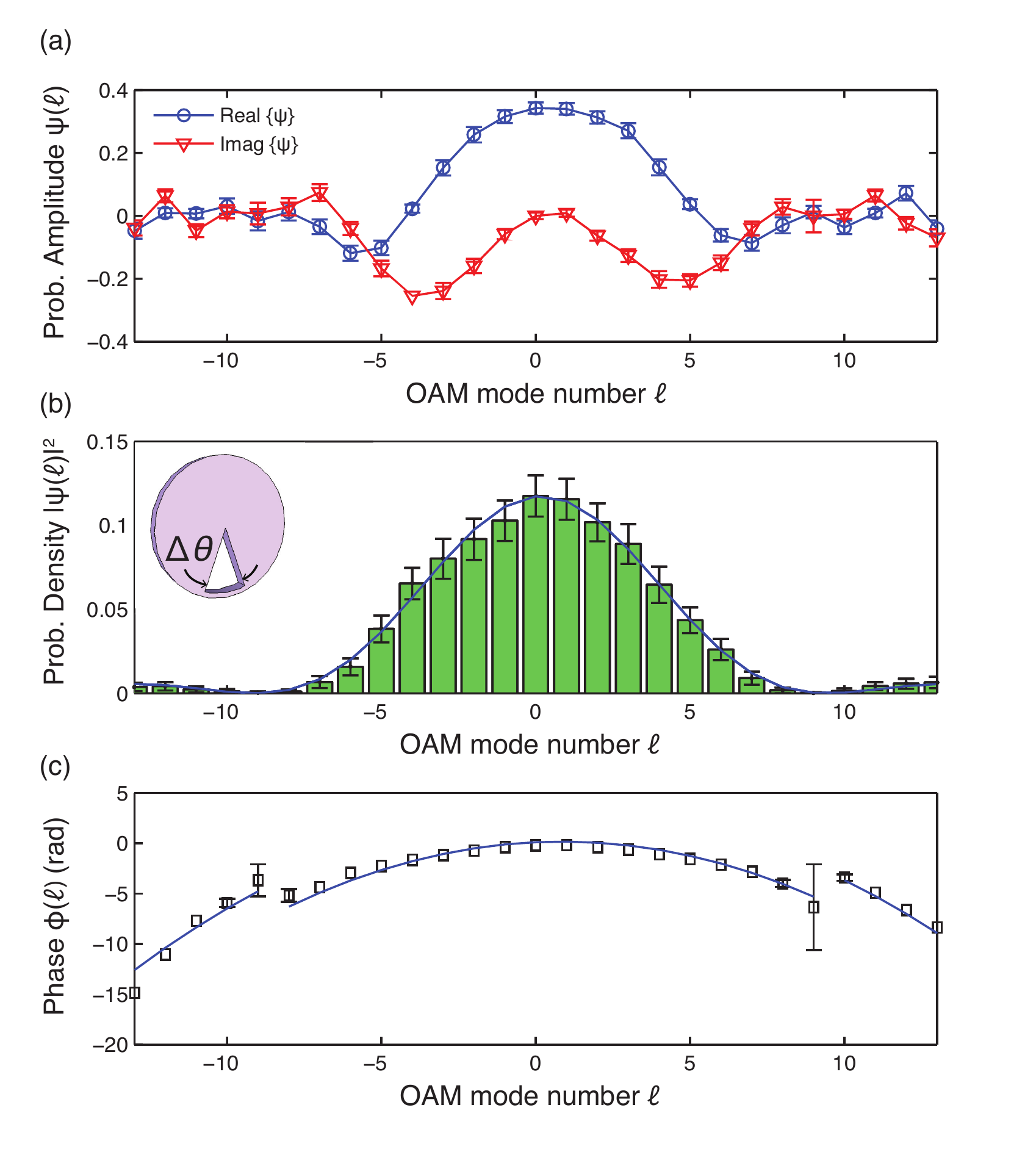}
\caption{Experimental data showing direct measurement of a 27-dimensional state vector in the OAM basis. The state is created by sending photons through an angular aperture of width $\Delta\theta = 2\pi/9$ rad (inset of (b)). (a) The measured real (blue circles) and imaginary parts (red triangles) of the state vector, (b) the calculated probability density $\abs{\Psi(\ell)}^2$, and (c) the calculated phase $\phi(\ell)$ are plotted as functions of the OAM quantum number $\ell$ up to a dimensionality of $\ell=\pm13$. The probability density has a sinc-squared shape. The phase has an asymmetric quadratic shape due to small misalignments in our optical system. Additionally, $\pi$-phase jumps are seen in the phase when the probability amplitude changes sign (not seen in the probability density). Theoretical fits to the probability density and phase are plotted as blue lines. Error bars are calculated by propagating the detector error (due to background light and dark counts) through to all measured quantities. Error bars larger than the symbols are shown. The data shown is the average result obtained from 50 experimental runs.}\label{results1}
\end{figure}  

%1315 words
\noindent\textbf{Direct measurement of a 27-dimensional OAM state.} The authors of the first work on direct measurement showed this technique to give identical results for heralded single photons and attenuated coherent states \cite{Lundeen:2011isa}. Therefore, in our experiment, photons from a highly attenuated HeNe laser are tailored into a high-dimensional quantum state by impressing a specific OAM distribution on them with SLM1 and a 4$f$ system of lenses (Fig. \ref{DMsetup}) \cite{Arrizon:2007wl}. The laser power is reduced such that probabilistically only one photon is present in our apparatus at any given time. First, we create a sinc-distribution of OAM using a wedge-shaped mask on the SLM. Just as a rectangular aperture diffracts light into a sinc-distribution of linear momenta, photons diffracting through an angular aperture of width $\Delta\theta$ result in a state vector with a sinc-distribution of OAM probability amplitudes \cite{Yao:2006vc} 

\begin{figure*}[t!]
\centering\includegraphics[scale=.6]{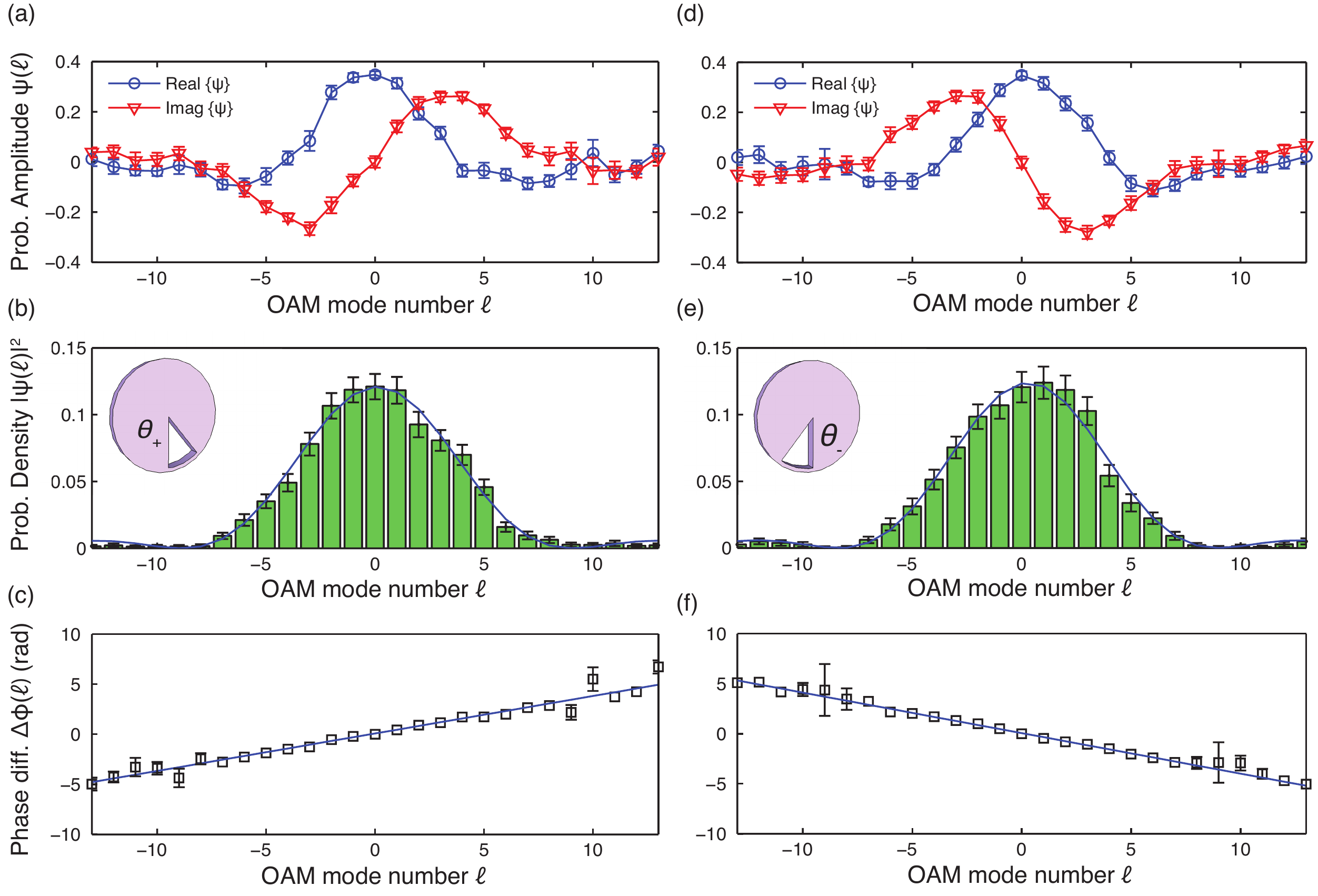}
\caption{Experimental data showing the direct measurement of a rotated high-dimensional state vector. The created state is rotated by angles $\theta_{\pm} = \pm\pi/9$ rad (insets of (b) and (e)). (a) and (d) The measured real (blue circles) and imaginary parts (red triangles) of the rotated state vectors. (b) and (e) The calculated probability densities $\abs{\Psi(\ell)_{\pm}}^2$. (c) and (f) The phase difference $\Delta\phi_{\pm}(\ell)$ between the calculated phase and the phase of the unrotated case from Fig. \ref{results1}(c). Theoretical fits to the probability densities and phases are plotted as blue lines. The linear fits in (c) and (f) are calculated via the process of chi-square minimization, which takes into account the error at each point. Error bars are calculated by propagating the detector error (due to background light and dark counts) through to all measured quantities. Error bars larger than the symbols are shown. The data shown is the average result obtained from 50 experimental runs.}\label{results2}
\end{figure*}

\be a_{\ell} = k\,\textrm{sinc}\bigg(\frac{\Delta\theta\ell}{2}\bigg).\ee

\noindent This distribution has a width given by $\Delta\ell=2\pi/\Delta\theta$, which refers to the mode index of its first null. Using an angular aperture of width $\Delta\theta=2\pi/9$ rad (inset of Fig. \ref{results1}(b)), we create such an ensemble of identical photons and perform the direct measurement procedure on them. The measured real and imaginary parts of the state vector are plotted in Fig. \ref{results1}(a) as a function of $\ell$. Using these quantities, we calculate the probability density $\abs{\Psi(\ell)}^2$ and the phase $\phi(\ell)$, which are plotted in Figs. \ref{results1}(b) and (c). The width of the sinc-squared fit to the probability density is measured to be $\Delta\ell=9.26\pm0.21$, which is very close to the value of $\Delta\ell=9$ predicted from theory. 

The measured phase plotted in Fig. \ref{results1}(c) has a quadratic form with $\pi$-phase jumps at OAM mode numbers $\ell=\pm 9$. These mode numbers correspond to the probability density minima in Fig. \ref{results1}(b), which is where the sinc-shaped amplitude crosses the x-axis and changes sign. The asymmetric quadratic feature in the phase appears due to small misalignments in our optical system. A $4f$ imaging system (not shown in Fig. \ref{DMsetup}) is used to magnify the Fourier plane of lens L2 onto SLM4. A misalignment in the $z$-axis of this imaging system appears as a quadratic phase. Further, the asymmetry in the phase is due to a first-order tilt aberration in our optical system, which is simply a result of the plane of SLM2 not being perfectly parallel to the plane of SLM4. Taking these two alignment imperfections into account, we use a quadratic model of the form $ax^2+bx+c$ in order to calculate a fit to the phase using a least-squares fitting algorithm in Matlab. Theoretical fits are plotted as blue lines in Fig. \ref{results1}(c). As can be seen, the phase error is unavoidably large when the amplitude approaches zero.\\

\noindent\textbf{Direct measurement of rotations in the OAM basis.} We now use this technique to analyze the effect of rotation on a photon carrying a broad range of angular momenta. Rotation of a state vector by an angle $\theta_0$ can be expressed by the unitary operator $\h{U}=\textrm{exp}(i\h{L}_z\theta_0)$, where $\h{L}_z$ is the angular momentum operator. Operating on our quantum state $\ket{\Psi}$ with $\h{U}$, we get

\be \ket{\Psi'} = \h{U}\ket{\Psi}=\sum_\ell k\,\textrm{sinc}\bigg(\frac{\Delta\theta\ell}{2}\bigg) e^{i\ell\theta_0}\ket{\ell}.\ee 

\noindent Thus, rotation by an angle $\theta_0$ manifests as an $\ell$-dependent phase $e^{i\ell\theta_0}$ in the OAM basis. For this reason, the angular momentum operator is called the generator of rotations under the paraxial approximation \cite{Barnett:1990do}. To measure this phase, we create a rotated state vector by rotating our angular aperture by an angle $\theta_+=\pi/9$ rad (inset of Fig. \ref{results2}(b)). Then, we perform the direct measurement procedure as before and measure the real and imaginary parts of the rotated state vector as a function of $\ell$ (Fig. \ref{results2}(a)). The probability density and phase of the state vector are calculated and plotted in Figs. \ref{results2}(b) and (c). For clarity, we subtract the phase of the zero rotation case (Fig. \ref{results1}(c)) from our phase reading, so the effect of rotation is clear. Barring experimental error, the amplitude does not change significantly from the unrotated case (Fig. \ref{results1}(b)). However, the phase of the OAM distribution exhibits a distinct $\ell$-dependent phase ramp with a slope of $0.373\pm0.007$ rad/mode. This is in close agreement with theory, which predicts the phase to have a form $\phi(\ell)=\pm\pi\ell/9$, corresponding to a phase ramp with a slope of $\pm 0.35$ rad/mode. A linear fit to the phase is calculated by the process of chi-square minimization, which takes into account the phase error at each point. This process is repeated for a negative rotation angle $\theta_-=-\pi/9$ rad, which results in a mostly unchanged probability density, but an $\ell$-dependent phase ramp as expected with a negative slope of $-0.404\pm0.007$ rad/mode (Figs. \ref{results2} (d)-(f)).

These results clearly illustrate the relationship between phase and rotation in the OAM basis in that every $\ell$-component acquires a phase proportional to the azimuthal quantum number $\ell$. The measured slopes in both cases are slightly larger than those expected from theory possibly due to errors introduced in the geometrical transformation that is used to spatially separate the OAM modes. The mode sorting process is extremely sensitive to choice of axis, and a very small displacement of the transforming elements R1 and R2 can propagate as a phase error.

%To further test our method, we rotate our state by an angle $\theta^-=-\pi/9$ rad (inset of Fig. \ref{results2}(d)), and plot the measurement results in Figs. \ref{results2}(d)-(f). The real part of the wavefunction (blue circles) does not change very much from the positive rotation case (Fig. \ref{results2}(a)). However, the imaginary part undergoes a distinct sign flip. This results in a mostly unchanged probability density, but an $\ell$-dependent phase ramp with a negative slope. The measured slope ($-0.41\pm??$ rad/mode) is a little different from that expected from theory ($-0.35$ rad/mode). This can be attributed to errors introduced in the geometrical transformation that is used to spatially separate the OAM modes. The mode sorting process is extremely sensitive to misalignment, and a very small displacement of the transforming elements can propagate as a phase error.

\section*{Discussion}

In conclusion, through weak measurements of orbital-angular-momentum (OAM) and strong measurements of angular position, we have measured the complex probability amplitudes that completely characterize a pure quantum state in the high-dimensional bases of OAM and angular position. Using our technique, we have also measured the effects of rotation on a 27-dimensional state vector in the OAM basis. The rotation manifests as an OAM mode-dependent phase and allows us to observe the action of the angular momentum operator as a generator of rotations \cite{Shankar:1994}. While we have directly measured pure states of OAM, this method can be extended to perform measurements of mixed, or general quantum states \cite{Lundeen:2012db,Salvail:2013bo}. By scanning the strong measurement of angular position as well, one can measure the Dirac distribution, which is informationally equivalent to the density matrix of a quantum state \cite{Dirac:1945wx,Chaturvedi:2006kz}. Further, photons entangled in OAM can be measured by extending this technique to two photons. In this case, one would need to perform independent weak and strong measurements on each photon, followed by a joint detection scheme for the polarization measurement. It is important to mention that while we have used a quantum description for the direct measurement method, it is perfectly explained using classical wave mechanics \cite{Bamber:tu}. However, the quantum mechanical description is simpler, more elegant, and extendable to systems that do not have a classical description.

Direct measurement may offer distinct advantages over conventional methods of quantum state characterization such as tomography. This method does not require a global reconstruction, a step that involves prohibitively long processing times for high-dimensional quantum states such as those of OAM \cite{Thew:2002tc,Agnew:2011js}. Consequently, the quantum state is more accessible in that it can be measured locally as a function of OAM quantum number $\ell$, as in our experiment. However, a detailed quantitative comparison between direct measurement and tomography will require careful consideration of parameters such as state reconstruction fidelity and will be the subject of future work. These advantages may allow us to use direct measurement for characterizing very large quantum states of OAM much more efficiently, with significant potential applications in high-dimensional quantum and classical communication systems \cite{Malik:2012ka,Boyd:2011ff,Mirhosseini:2013go}. Furthermore, the direct measurement procedure is not just limited to optical systems such as ours and can be used for the characterization of other high-dimensional quantum systems.

\section*{Methods}

\textbf{Experimental details.} Elements R1 and R2 are freeform refractive elements made out of PMMA that map polar coordinates $(r,\phi)$ to rectilinear coordinates $(x,y)$ through the log-polar mapping $x=a(\phi\,\textrm{mod}\,2\pi)$ and $y=-a\ln{(r/b)}$ \cite{Berkhout:2010cb}. These are used for transforming OAM modes with azimuthal phase variation $\exp(i\ell\phi)$ to plane wave modes with position phase variation $\exp(i\ell x/a)$. R1 unwraps the phase and R2 removes a residual aberration introduced in the unwrapping process. These elements were machined using a Nanotech ultra precision lathe in combination with a Nanotech NFTS6000 fast tool servo. The optical thickness of element R1 can be written as a function of $(x,y)$ as \cite{Lavery:2012hi}

\bea Z_1(x,y) =& \frac{a}{f(n-1)}\bigg[y\,\textrm{arctan}(y/x)-x\ln(\sqrt(x^2+y^2)/b)\nonumber\\
&+x-\frac{1}{2a}(x^2+y^2)\bigg].\eea 

\noindent where $f$ is the focal length of the lens integrated into both elements. This lens performs the Fourier transform operation that is required between the unwrapping and phase correcting procedures \cite{Berkhout:2010cb}. The two free parameters, $a$ and $b$, dictate the size and position of the transformed beam. The optical thickness of element R2 can be similarly written as

\bea Z_2(x,y) = -\frac{ab}{f(n-1)}\bigg[\exp\bigg(-\frac{u}{a}\bigg)\cos\bigg(\frac{v}{a}\bigg)\nonumber\\
-\frac{1}{2ab}(u^2+v^2)\bigg].\eea

\noindent where $u$ and $v$ are spatial cartesian coordinates in the output plane. The distance between these two elements must be exactly $f$, and the elements must be aligned precisely along the same optical axis. For this reason, they are mounted in a cage system with fine position and rotation controls. 

After element R2, the component OAM modes of the photon still have an overlap of about 20\%. This is due to the finite size of the transformed momentum mode, which is bounded by the function $\textrm{rect}(x/2\pi a)$. A fan-out hologram \cite{Prongue:1992uj,OSullivan:2012gj} implemented on a phase-only spatial light modulator (SLM2) creates three adjacent copies of the momentum mode. The fan-out hologram is calculated from values given in Refs. \cite{Prongue:1992uj,Romero:2007vi}. The three copies generated by SLM2 have a phase offset from one another, which is removed by a phase-correcting hologram implemented on SLM3. The larger momentum modes created by this process result in smaller position modes when Fourier transformed by a lens.
%
%\begin{figure}[t!]
%\centering\includegraphics[scale=0.5]{Figures/fanoutX.eps}
%  \caption{One-dimensional phase profiles of the (a) fan-out hologram used for creating 3 copies and (b) phase correcting hologram used to remove the phase offset between the 3 copies. The phase profile in (a) shows a small section of the SLM.}\label{fanoutX}
%\end{figure}

SLM1, SLM2 and SLM4 are Holoeye PLUTO phase-only SLMs antireflection coated for visible light. These SLMs have a spatial resolution of 1920x1080 pixels and a pixel size of 8 $\mu$m. These are used for mode generation \cite{Arrizon:2007wl}, beam copying (fan-out) \cite{Prongue:1992uj}, and polarization rotation \cite{Davis:2000ur}. SLM3 is a Cambridge Correlator SDE1024 liquid-crystal-on-silicon SLM with a resolution of 1024x768 pixels. This SLM is used as the phase-corrector for the fan-out hologram, as the lower resolution is sufficient for this process. The quarter and half-wave plates used in our experiment are zero-order wave plates manufactured by Thorlabs and optimized for a wavelength of 633 nm. The slit used for post-selection is 10 $\mu$m wide and is also manufactured by Thorlabs. The single photon avalanche detectors (SPADs) are Perkin Elmer SPCM-AQRH-14 modules with a dark count rate of 100 counts/second. 
\\\\
\noindent\textbf{OAM weak value derivation.} Here we derive the relationship between the OAM weak value $\avg{\pi_{\ell}}_{\textrm{w}}$ and expectation values of the $\h\sigma_1$ and $\h\sigma_2$ Pauli operators (Eq. (\ref{Pauli}) in the text). The von Neumann formulation can be used to describe the coupling between the OAM (system) and polarization (pointer) observables \cite{Neumann:1955,Lundeen:ta}. The product Hamilton describing this interaction can be written as

\be\h{H}=-g\,\h{\pi}_{\ell}\cdot\h{S}_2=-\bigg(\frac{g\,\hbar}{2}\bigg)\,\h{\pi}_{\ell}\cdot\h\sigma_2,\ee

\noindent where $g$ is a constant indicating the strength of the coupling, $\h{\pi}_{\ell}$ is the projection operator in the OAM basis, and $\h\sigma_2$ is the Pauli spin operator in the $y$ direction on the Bloch sphere (not to be confused with the coordinate space describing the polarization). The measurement pointer is initially in a vertical polarization state

\be \ket{s_\textrm{i}} = \left[ \begin{array}{c}
0 \\
1 \\
\end{array} \right] \label{initial}\ee 

\noindent and the system is in an initial state $\ket{I}$. The initial system-pointer state is modified by a unitary interaction $\h{U}=\exp(-i\h{H}t/\hbar)$, which can be written using the product Hamiltonian above as

\be\h{U}=\exp\bigg(\frac{i\,gt\,\h{\pi}_{\ell}\cdot\h\sigma_2}{2}\bigg)=\exp\bigg(\frac{i\,\sin\alpha\,\h{\pi}_{\ell}\cdot\h\sigma_2}{2}\bigg)\ee

\noindent Here we have substituted $\sin\alpha$ in place of $gt$ as a coupling constant. This refers to the angle $\alpha$ by which we rotate the polarization of the OAM mode to be measured in our experiment. When $\alpha$ is small, the measurement is weak. In this case, we can express the operator $\h{U}$ as a Taylor series expansion truncated to first order in $\sin\alpha$. The initial state then evolves to

\bea \ket{\Psi(t)} &=& (1-\frac{i\h{H}t}{\hbar}-...)\ket{I}\ket{s_\textrm{i}}\nonumber\\
&=& \ket{I}\ket{s_\textrm{i}}+\frac{i\,\sin\alpha}{2}\h{\pi}_{\ell}\ket{I}\h\sigma_2\ket{s_\textrm{i}}
\eea

\noindent We can express the strong measurement as a projection into a final state $\ket{F}$:

\be \bra{F}\h{U}\ket{I}\ket{s_\textrm{i}}=\brac{F}{I}\ket{s_\textrm{i}}+\frac{i\,\sin\alpha}{2}\bra{F}\h{\pi}_{\ell}\ket{I}\h\sigma_2\ket{s_\textrm{i}}\ee

\noindent We can then divide by $\brac{F}{I}$ to get the final pointer polarization state:

\bea \ket{s_\textrm{f}} &=& \ket{s_\textrm{i}}+\frac{i\,\sin\alpha}{2}\frac{\bra{F}\h{\pi}_{\ell}\ket{I}}{\brac{F}{I}}\h\sigma_2\ket{s_\textrm{i}}\nonumber\\
&=& \ket{s_\textrm{i}}+\frac{i\,\sin\alpha}{2}\avg{\pi_{\ell}}_{\textrm{w}}\h\sigma_2\ket{s_\textrm{i}}
\eea

\noindent Notice that the weak value $\avg{\pi_{\ell}}_{\textrm{w}}=\bra{F}\h{\pi}_{\ell}\ket{I}/\brac{F}{I}$ appears in the above equation. Using this expression for the final state of the pointer, we can calculate the expectation value of $\h\sigma_1$ as follows:

\bea \bra{s_\textrm{f}}\h\sigma_1\ket{s_\textrm{f}} =& \cancel{{\bra{s_\textrm{i}}\h\sigma_1\ket{s_\textrm{i}}}} +\frac{i\,\sin\alpha}{2}\bigg[\avg{\pi_{\ell}}_{\textrm{w}}\bra{s_\textrm{i}}\h\sigma_1\h\sigma_2\ket{s_\textrm{i}}\nonumber\\
&-\avg{\pi_{\ell}}_{\textrm{w}}^\dag\bra{s_\textrm{i}}\h\sigma_2\h\sigma_1\ket{s_\textrm{i}}\bigg]
\eea

\noindent Using the substitution $\avg{\pi_{\ell}}_{\textrm{w}}$ = Re\{$\avg{\pi_{\ell}}_{\textrm{w}}$\} + $i$Im\{$\avg{\pi_{\ell}}_{\textrm{w}}$\} and the initial state $\ket{s_\textrm{i}}$ from Eq. (\ref{initial}), the above equation can simplified further:

\bea \bra{s_\textrm{f}}\h\sigma_1\ket{s_\textrm{f}} &=&  \frac{i\,\sin\alpha}{2}\bigg[\textrm{Re}\{\avg{\pi_{\ell}}_{\textrm{w}}\}\bra{s_\textrm{i}}\h\sigma_1\h\sigma_2-\h\sigma_2\h\sigma_1\ket{s_\textrm{i}}\nonumber\\
&&+ \;i\,\textrm{Im}\{\avg{\pi_{\ell}}_{\textrm{w}}\}\bra{s_\textrm{i}}\cancel{{\h\sigma_1\h\sigma_2+\h\sigma_2\h\sigma_1}}\ket{s_\textrm{i}}\bigg]\nonumber\\
&=& -\sin\alpha\,\textrm{Re}\{\avg{\pi_{\ell}}_{\textrm{w}}\}\bra{s_\textrm{i}}\h{\sigma}_3\ket{s_\textrm{i}}\nonumber\\
&=& \sin\alpha\,\textrm{Re}\{\avg{\pi_{\ell}}_{\textrm{w}}\}
\eea

\noindent Similarly, we can calculate the expectation value of $\h\sigma_2$ as follows:

\bea 
\bra{s_\textrm{f}}\h\sigma_2\ket{s_\textrm{f}} &=& \cancel{{\bra{s_\textrm{i}}\h\sigma_2\ket{s_\textrm{i}}}} +\frac{i\,\sin\alpha}{2}\bigg[\avg{\pi_{\ell}}_{\textrm{w}}\bra{s_\textrm{i}}\h\sigma_2\h\sigma_2\ket{s_\textrm{i}}\nonumber\\
&&-\avg{\pi_{\ell}}_{\textrm{w}}^\dag\bra{s_\textrm{i}}\h\sigma_2\h\sigma_2\ket{s_\textrm{i}}\bigg]\nonumber\\
&=&  \frac{i\,\sin\alpha}{2}\bigg[\textrm{Re}\{\avg{\pi_{\ell}}_{\textrm{w}}\}\bra{s_\textrm{i}}\cancel{{\h\sigma_2\h\sigma_2-\h\sigma_2\h\sigma_2}}\ket{s_\textrm{i}}\nonumber\\
&&+\; i\,\textrm{Im}\{\avg{\pi_{\ell}}_{\textrm{w}}\}\bra{s_\textrm{i}}\h\sigma_2\h\sigma_2+\h\sigma_2\h\sigma_2\ket{s_\textrm{i}}\bigg]\nonumber\\
&=& -\sin\alpha\,\textrm{Im}\{\avg{\pi_{\ell}}_{\textrm{w}}\}\bra{s_\textrm{i}}\h\sigma_2^2\ket{s_\textrm{i}}\nonumber\\
&=& -\sin\alpha\,\textrm{Im}\{\avg{\pi_{\ell}}_{\textrm{w}}\}
\eea

\noindent Thus, we see that the real and imaginary parts of the OAM weak value $\avg{\pi_{\ell}}_{\textrm{w}}$ are proportional to the expectation values of the $\h\sigma_1$ and $\h\sigma_2$ Pauli operators (Eq. (\ref{Pauli}) in the text):

\bea \avg{\pi_{\ell}}_{\textrm{w}} &=& \textrm{Re}\{\avg{\pi_{\ell}}_{\textrm{w}}\} + i\,\textrm{Im}\{\avg{\pi_{\ell}}_{\textrm{w}}\}\nonumber\\
&=& \frac{1}{\sin\alpha}\big[\bra{s_\textrm{f}}\h\sigma_1\ket{s_\textrm{f}}-i\bra{s_\textrm{f}}\h\sigma_2\ket{s_\textrm{f}}\big]
\eea

\section*{Acknowledgements}

This work was supported by the DARPA InPho program, the Canadian Excellence Research Chair (CERC) program, the Engineering and Physical Sciences Research Council (EPSRC), the Royal Society, and the European Commission through a Marie Curie fellowship. M.Malik would like to thank Dr. Justin Dressel for helpful discussions.

\section*{Author Contributions}

M.Malik devised the concept of the experiment. M.Malik and M.Mirhosseini performed the experiment and analyzed data. M.P.J.L. assisted with the experiment. J.L. advised on early aspects of experimental design. R.W.B. and M.J.P. supervised the project. M.Malik wrote the manuscript with contributions from all authors.

%\bibliography{DM_OAM,DMcustom}

%\bibliographystyle{apsrev4-1}

\end{document}